\documentstyle[epsfig]{article}
\begin{document}
\title{Quantum coherence generated by
  interference-induced state selectiveness}
\author{Jean Claude Garreau \\
Laboratoire de Physique des Lasers, Atomes et Mol{\'e}cules and\\
Centre d'Etudes et de Recherches Laser et Applications,\\
Universit\'{e} des Sciences et Technologies de Lille, \\
F-59655 Villeneuve d'Ascq Cedex, France\\}
\maketitle

\begin{abstract}
  The relations between quantum coherence and quantum interference are
  discussed. A general method for generation of quantum coherence through
  interference-induced state selection is introduced and then applied to 
  `simple' atomic systems under two-photon 
  transitions, with applications in quantum optics and
  laser cooling.
\end{abstract}

\section{Introduction}
\label{sec:Intro}
Quantum mechanics, as it is described by the Schr\"odinger's equation,
presents the striking feature of imposing a statistical description even
for a lone particle. However, when one deals with
large amounts
of quantum particles, as is the case in most experiments,
this fundamental quantum aspect is generally washed out.
Statistics of identical quantum particles is still different
from classical statistics: 
identical particles behave as bosons or fermions, i.e. they
do not obey Boltzmann's law. Still, if
one deals with complex systems, like atoms, having a large number of
eigenstates,
at temperatures that are large compared to the typical
energy interval between the eigenstates (i.e. for low occupation
numbers) this aspect is also erased, as both
Bose-Einstein and Fermi-Dirac statistics tend to Boltzmann's law.
Even then quantum behavior
can differ from classical behavior, because the amplitudes of probability
appearing in Schr\"odinger equation are complex, and thus have a phase:
{\em Quantum amplitudes can interfere}.

The global phase of a quantum state cannot be measured directly 
(otherwise quantum mechanics would not be invariant under a change of the 
zero of energy), but {\em relative} phases can, and are, observed in quantum 
interference experiments. The linearity of Schr\"odinger's equation 
insures that linear superpositions of solutions are also solutions.
Probabilities calculated for such superpositions are
sensitive the relative phase of the states forming the
superposition, and produce quantum interference patterns.
The name `coherent state' is thus 
justified for such states as well as
the name of `quantum coherence' for the associated
property. Formally, the quantum coherence 
of a system corresponds to the non-diagonal elements of the density
operator. This definition can be extended to systems
mixing quantum and classical statistics, and puts into evidence
the fact that coherence is a base-dependent concept.

Quantum coherence is a {\it sine qua non} condition for the observation of
quantum interference, which in turn is one of the most common ways
of generating 
`nonclassical' effects, like light-noise squeezing or quantum beats.
Interestingly
enough, quantum coherence is also essential in producing ``classical" states
like the coherent states of the electromagnetic field.

In the present paper we shall somewhat turn the quantum
interference problem upside
down: can quantum interference generate `non-trivial' types
of quantum
coherence? After describing a general method for generating quantum
coherence by using quantum interference, we shall illustrate this main
idea with a few simple examples. We
shall concentrate the discussion on the physical mechanisms, detailed
quantitative approaches can be found in the references.

\section{Generating quantum coherence with quantum interference}
\label{sec:method}

\begin{figure}
\begin{center}
  \epsfig{figure=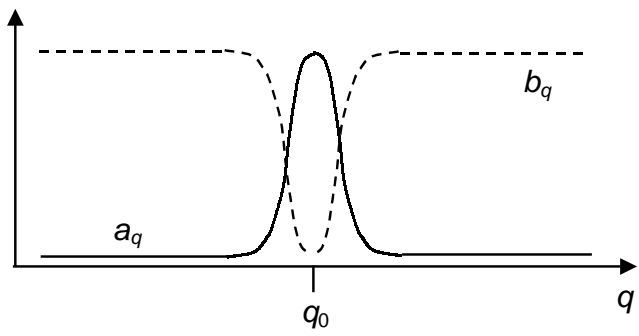,width=12cm,clip=}
\end{center}
\caption{Typical probability distributions for long evolution times
  generated by selection through quantum interference.}
\label{fig:Distributions}
\end{figure}

The method we consider here is discussed in detail in
ref. \cite{ref:Cohint}. The generation
of coherence relies on the
state selectiveness of the quantum
interference. In order to illustrate the principle, consider a system
with two degrees of freedom. The first (`internal')
degree of freedom corresponds to
two levels $|g \rangle$ and $|s \rangle$, whereas the second
one is associated with states $|q \rangle$ that can be discrete or
continuous. We suppose that a perturbation couples
the states $|g,q_g\rangle$ to the states $|s,q_s\rangle$ in an
irreversible way, and that the relative phase of such states depends
on the parameter $q$, so that the quantum interference can be
`tuned' to produce complete destructive
interference for some value $q_g=q_0$. If the initial state of the
system is of the form $\sum_q a_q(0) |g,q\rangle$, under the action
of the coupling perturbation the terms in
the sum with $q \neq q_0$ will progressively
perform transitions to
the corresponding states $|s,q\rangle$, so that the overall state
of the system after a time $t$ has elapsed is:
\begin{equation}
\label{eq:generalstate}
\sum_q \left[ a_q(t) |g, q\rangle + b_q(t) |s,q\rangle \right] \mathrm{.}
\end{equation}
As $t$ increases, $a_q(t)$ becomes more and more concentrated around
$q=q_0$, whereas $b_q(t)$ will present a more and more pronounced
`hole' around $q=q_0$, as shown
schematically in Fig.~\ref{fig:Distributions}. The final state shall
thus asymptotically tend to
\begin{equation}
a_{q_0} |g,q_0\rangle +
\sum_{q \neq q_0} b_q |s,q\rangle \mathrm{.}
\label{eq:asymptstate}
\end{equation}
Clearly, the evolution due to the perturbation has generated
an {\em entangled} state of the two degrees of freedom by
{\em selection} of the $q=q_0$ value.
Suppose one performs at large enough time $t$ a measurement
of the {\em internal} state corresponding to the first degree
of freedom and founds
$|g \rangle$. The system is then projected onto the state
$\sum_q a_q(t)|g,q\rangle$, which means that
the probability of finding the system in the state $|q_0\rangle$
is very close to one (cf. Fig.~\ref{fig:Distributions}).
However, the probability of getting the state
$|g\rangle$ as the result of the measurement is
$\sum_q |a_q(t)|^2$ which tends to
zero as $t$ increases. This `depopulation' effect is the main
limitation of the present method: the sharper the
selection, the smaller the probability of getting the `good'
state selection. We shall consider in section \ref{sec:cooling}
methods for controlling the depopulation of the selected states.

In discussing specific examples in the following
sections, we shall work essentially with the `diamond' level scheme
shown in Fig.~\ref{fig:levels}. 
A quantum object called `the atom' has four `internal' states.
States $|g\rangle$ and $|s\rangle$ are coupled by a two-photon transition via
two different intermediate states $|e_{1,2}\rangle$. A mode of the
electromagnetic field labeled 1 (resp. 2) couples the ground state to
the intermediate state $|e_1\rangle$ (resp. $|e_2\rangle$),
and the intermediate state $|e_2\rangle$ (resp. $|e_1\rangle$) to
the excited state $|s\rangle$. Whether the excited state
is a discrete state, a band, a continuum, or even a discrete state
coupled itself to a continuum is not important here.

The excited state is connected to the ground state
by two indistinguishable
paths, which leads to quantum interference. Using second order perturbation,
the transition rate can be written as:
\begin{equation}
\label{eq:transrate}
W = { n_1 n_2 \over \hbar^2 } \left| { A_1 \over \delta_1^\prime } 
        + { A_2 \over \delta_2^\prime } \right|^2
\end{equation}
where $\delta_i^\prime$ is the {\em effective} detuning of
the intermediate levels, which might take into account Doppler effect
and/or light-shift effects, and $n_{1,2}$ are the photon number in each mode.
The most general expression for the detuning we shall consider in
the following is
\begin{equation}
\label{eq:effectivedet}
\delta_i^\prime = \delta_i + \beta_{ii} n_i + \beta_{ij} n_j + k_i v
\end{equation}
where $\beta_{ij}$ are light-shift coefficients (that can be easily calculated
from perturbation theory), $k_i$ is the wavenumber of mode $i$,
and $v$ the center of mass velocity. The light-shift and the Doppler
effect play here the role of a `coupling' between degrees of freedom
linking the {\em internal}
atomic state respectively
to the electromagnetic field and to the center of mass motion.

Eq.\ (\ref{eq:transrate}) vanishes if
\begin{equation}
\label{eq:interfcondition}
A_1 \delta_2^\prime = - A_2 \delta_1^\prime \mathrm{ .}
\end{equation}
The system parameters ($n_1, n_2$ or $v$)
appear in the expression of the detunings $\delta_i^\prime$.
The above interference condition can thus be satisfied only for
particular values of such parameters, and the corresponding state
will be selected, as discussed above.
Note however that it is not always possible to satisfy the
interference condition, for example if the amplitudes $A_{1,2}$
have different complex phases. This is the principle of the method
of quantum coherence generation we shall apply to specific examples
in the following of the paper.

\section{Generation of two-mode coherence}
\label{sec:lightcoherence}

Let us now consider the case in which the light-shift
coupling between the internal state and the field modes creates
quantum coherence between the latter. We
neglect the Doppler effect compared to the light-shifts in
Eq.\ (\ref{eq:effectivedet}). The destructive interference condition
deduced from Eqs.\ (\ref{eq:effectivedet}) and
(\ref{eq:interfcondition})
is:

\begin{equation}
  n_2 = \alpha n_1 + \mu
  \label{eq:phnbselect}
\end{equation}
where
\begin{equation}
\alpha = - { A_1 \beta_{22} + A_2 \beta_{12} \over
  A_1 \beta_{21} + A_2 \beta_{22} }
  \label{eq:alpha}
\end{equation}
\begin{equation}
\label{eq:mu}
\mu = - { A_1 \delta_2 + A_2 \delta_1 \over
  A_1 \beta_{21} + A_2 \beta_{22} } \mathrm{.}
\end{equation}
The condition for applying the above equations is
that $n_2 \ge 0$, which means that the detunings must be chosen
in the appropriate way (as the amplitudes
$A_i$, the detunings $\delta_i$ and the light-shift
coefficients $\beta_{ij}$ can be negative, this
condition is, in general, not too hard to satisfy). 

In order to simplify the notation, we take $A_1 = -A_2 \equiv A$.
We suppose that the modes are initially ($t=t_0$) in uncorrelated
states. The initial state of the system can thus be
written in the quite general form:
\begin{equation}
  |\psi_0\rangle = \sum_{n_1,n_2} a(n_1,n_2) |g; n_1,n_2 \rangle
  \label{eq:initialstate}
\end{equation}
Because of the coupling with the level $|s\rangle$, the population
of each term in the above initial state decays with a ratio given
by Eq. (\ref{eq:transrate}). If we consider the system's wave
function $|\psi(t)\rangle$ at a latter time $t_f$, and project
it onto the internal state $|g\rangle$ we obtain:
\begin{equation}
  \langle g|\psi(t_f)\rangle =
  \sum_{n_1,n_2} \exp[-W(n_1,n_2)t] a(n_1,n_2) |g; n_1,n_2 \rangle
  \label{eq:finalstate}
\end{equation}
Clearly, for sufficiently large values of $t_f$, only the terms
for which the interference condition Eq. (\ref{eq:phnbselect}) is
satisfied, such that $W(n_1,n_2)t \ll 1$ will survive. Quantum
interference has performed state selectiveness, and the
internal ground state $|g\rangle$ has become {\em entangled}
with a particular combination of photon numbers in modes 1 and 2.

\begin{figure}
\begin{center}
  \epsfig{figure=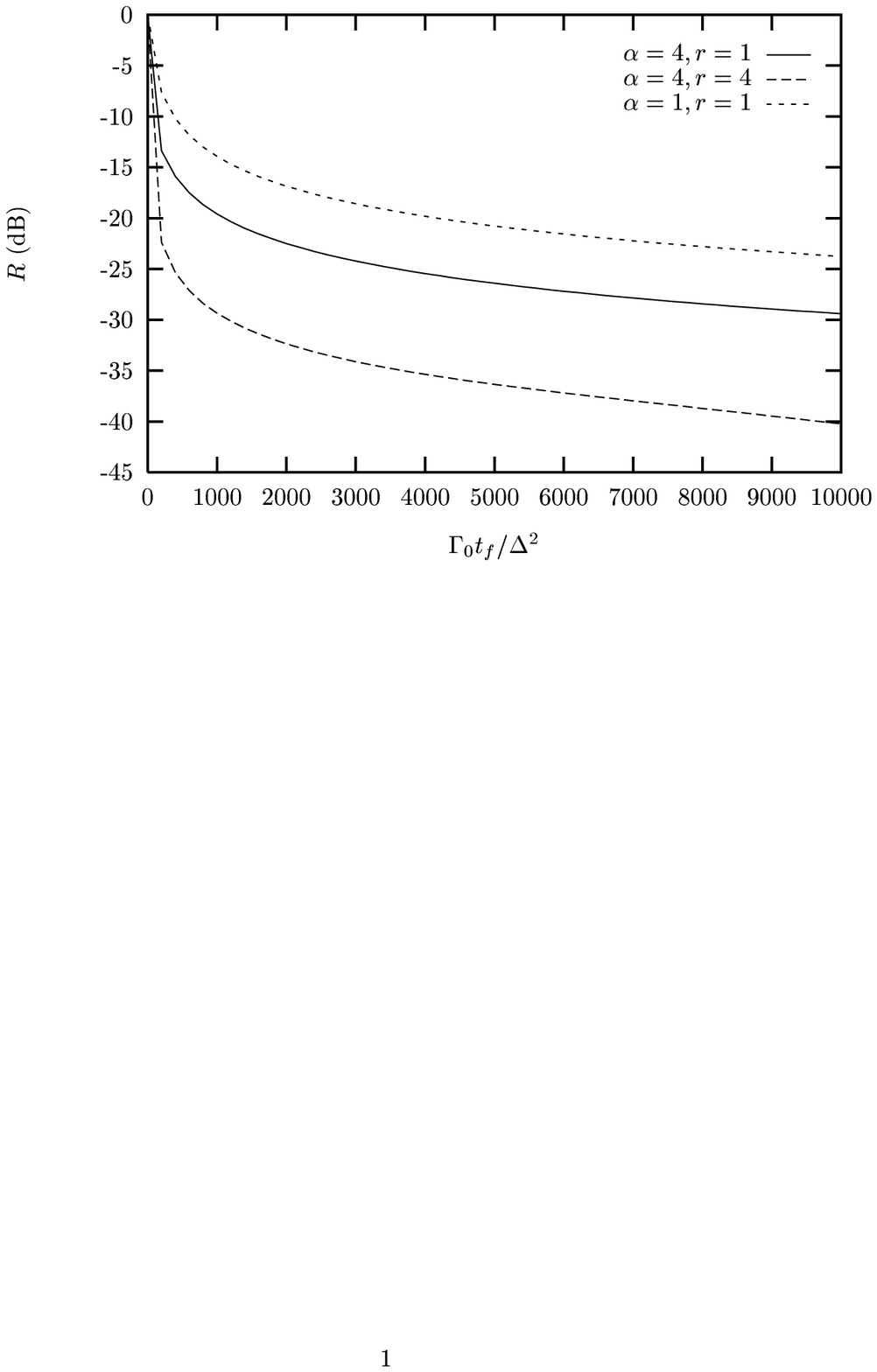,width=12cm,clip=}
\end{center}
\caption{Time evolution of the correlation between the two modes
of the electromagnetic field. The quantity $\Gamma_0$ is the normalized
ionization rate $|A|^2/\hbar \omega_0$. Other parameters are 
$\langle n_1 \rangle = 100$, $\beta_{11} = \beta_{22} = 10^{-4} |\delta|$,
$\beta_{12} = \beta_{21} =2 \times 10^{-5} |\delta|$ ($\delta$
is the detuning of the one-photon transition).}
\label{fig:degreeofcorr}
\end{figure}

In order to quantify this entanglement, consider the quantity
$\delta n \equiv n_2 - (\alpha n_1 + \mu)$. The fluctuations
of $\delta n$, $\langle \delta n^2 \rangle - \langle \delta n \rangle^2$,
constitute a measurement of the degree of correlation between the two
modes. If the two mode are independent, the fluctuation of $\delta
n$ is the sum of the fluctuations of each mode. As the modes become
correlated, this quantity should go to zero. Let us thus
suppose that at time $t_f$ one performs a measurement of the atomic state;
if the result is $|g\rangle$ we measure the degree of coherence
defined by the quantity:
\begin{equation}
\label{eq:degofcoherence}
R = 10 \log_{10} \left( { \langle \delta n^2(t_f) \rangle -
                          \langle \delta n(t_f) \rangle^2 \over
                          \langle \delta n^2(t_0) \rangle -
                          \langle \delta n(t_0) \rangle^2 } \right)
\mathrm{ .}
\end{equation}
The smaller $R$, the greater the degree of correlation. Physically,
one can interpret $R$ as the quantum noise reduction in the measurement of
the quantity $\delta n$ with respect to the `standard limit' 
corresponding to uncorrelated modes.
Fig.~\ref{fig:degreeofcorr} shows the time evolution of the quantity $R$
for three combinations of $(\alpha,r)$ where
$r \equiv \langle n_2\rangle/\langle n_1 \rangle$ is the ratio of
the mean photon numbers in modes 2 and 1. We took $\mu=0$.
The two modes are initially in coherent states.
Some interesting features of this result can be understood in terms
of rather simple arguments.

The first observation is that the degree of coherence increases
when $\alpha$ increases. As
the initial coherent state contains
the vacuum state, there is a finite probability of finding zero
photons in one of the field modes. Zero photons obviously means no
transition and consequently no generation of coherence: The 
coherence is so to say `contaminated' by the vacuum state. 
The probability of
this `vacuum tail' for coherent states decreases exponentially with
the mean photon number, so that higher values of $\alpha$
correspond to higher degrees of coherence.

A second interesting feature of the result is that the degree
of coherence is maximum for $r =\alpha$. This is a consequence of the
${\em filtering}$ character of the process:
The weight of the filtered state in the final overall
state is proportional to the weight of the filtered
state in the
{\em initial} overall state. This means that the final population of
the state $|n_1, \alpha n_1 \rangle$ in the final state is
proportional to the weights of the states $|n_1\rangle$ and
$|n_2=\alpha n_1\rangle$ in the {\em initial} state, that are maxima
if $\langle n_2 \rangle \approx \alpha \langle n_1 \rangle$
\cite{ref:Cohint}.

Let us finally mention that the present method can generate
`unusual' types of
correlations between the field modes, as the following examples
show. For $\alpha = 1$
and $\mu = 0$ one generates the usual kind of correlation obtained
in parametric down conversion (that is, $n_1 = n_2$). If $\alpha < 0$ and 
$\mu \geq |\alpha|$ one should have
$n_1 \leq \mu/|\alpha|$ and $n_2 \leq \mu$, and the resulting state
is a {\em finite} superposition of number states in each mode. In
particular, for $\mu=|\alpha|=1$ one should generate a one-photon
number state in each mode. If
$\alpha n_1 \ll \mu$ the final state tends to the number state $n_2=\mu$.

If $\alpha \gg \mu$, mode
2 becomes an `amplified replica' of mode 1: the quantum properties
of mode 1 have been `cloned' to mode 2, including quantum noise
properties \cite{ref:QCA}.

Closing this section, let us mention that experimental 
evidence of coherence generation in systems very similar to
the present one has been presented by Wang, Chen and Elliot
\cite{ref:expcohint}.

\section{Atomic velocity selection}
\label{sec:cooling}

Laser cooling of neutral atoms has been a main issue
of atomic physics for more then a decade now. 
Well understood mechanisms like Sisyphus cooling currently
allows to achieve 
temperatures close to the so-called `recoil limit'. This limit
corresponds to the fact that, if the {\em average}
momentum transmitted to the atom by spontaneously emitted
photons is zero,
the root mean square value of this momentum shift
is equal to the momentum $\hbar k$ of the photon. This sets
a minimum energy achievable by any cooling method involving spontaneous 
emission to $E_r = \hbar^2 k^2/2M$ (M is the mass of the atom), 
called `recoil energy', to which
one associates the `recoil temperature' $T_r = \hbar^2 k^2/M k_B$
($k_B$ is the Boltzmann's constant).
Nevertheless, it has been demonstrated that
by preventing low-velocity atoms from absorbing
photons it is possible achieve `subrecoil' temperatures.
Two such methods have been 
experimentally demonstrated: VSCPT (Velocity Selection
by Coherent Population Trapping) \cite{ref:VSCPT}
and Raman cooling \cite{ref:Raman}. VSCPT relies on the
possibility of generating ground-state coherences which, for
particular values of the atomic momentum, are uncoupled to the
electromagnetic radiation due to a quantum interference
effect \cite{note:EIT}. Once the atom attains a low enough momentum,
it is trapped in this momentum-defined coherent superposition and do
not undergo further fluorescence cycles.
We shall discuss Raman cooling in more detail in the next
section.

The general method described in Sec. \ref{sec:method}
can be applied to the
generation of coherence between the internal and external (center
of mass motion) degrees of freedom of an atom. By performing
a measurement of the internal state, 
very sharp atomic velocity selection is possible. The 
coupling between the degree of freedom is provided by the Doppler effect.
The process does not
involve spontaneous emission and in principle is not limited
by the recoil energy.

With counterpropagating laser beams of the same frequency, the 
destructive interference condition can be tuned to the
velocity class $v$ by choosing (we neglect light-shift terms)
\begin{equation}
\label{eq:interfcond}
v = { A_1 \delta_2 + A_2 \delta_1 \over (A_2-A_1) k }
\end{equation}
where we supposed that $A_1 \neq A_2$ (otherwise
destructive interference can not be achieved).

Paralleling the reasoning of the preceding section, we can
write the initial wave function of the system as
\begin{equation}
  |\psi_0\rangle = \int dv \, f(v) \, |g;v> \mathrm{.}
\end{equation}
The transition rate $W(v)$ now depends on the velocity $v$
via the detunings $\delta_{1,2}^\prime = \delta_{1,2} \pm kv$,
so that the projection of the asymptotic wave function over the
internal state $|g\rangle$ can be written:
\begin{equation}
  |\psi(t)\rangle = \int dv \, f(v) \, \exp[-W(v)t_f] \, |g;v> \mathrm{,}
\end{equation}
and, again, only velocity states such that $W(v)t_f \ll 1$ 
survive.

\begin{figure}
\begin{center}
  \epsfig{figure=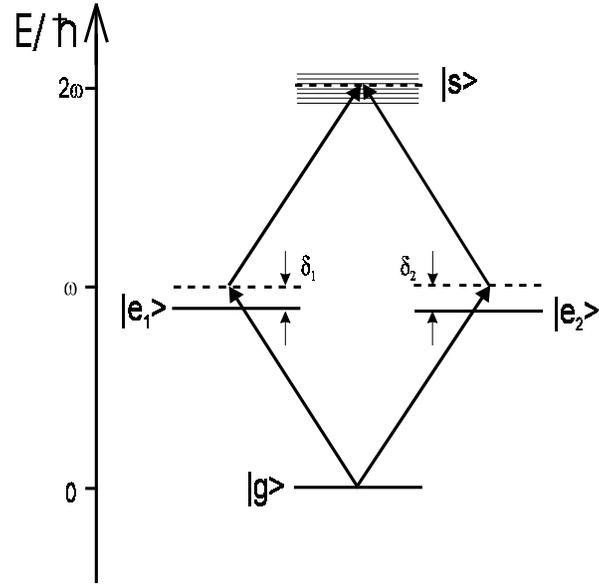,width=8cm,clip=}
\end{center}
\caption{`Diamond' level scheme for generation of quantum coherence by
quantum interference. Two modes of the electromagnetic field couple the
ground state $|g\rangle$ to the excited states $|s\rangle$ (that can form
continuum) via two 
intermediate states. The two paths leading to the excited level produce 
quantum interference depending on the parameters of the ground state
(e. g. center of mass velocity) and of the field modes (e.g. photon number).}
\label{fig:levels}
\end{figure}

The present case presents however a complication.
Referring to Fig.~\ref{fig:levels}, one sees that one
must also take into account the `stray' transition
in which the atom absorbs one photon, goes to one of the
intermediate levels and 
then decays back to the ground state by spontaneous emission.
With a negative detuning, this process is the usual Doppler cooling 
whose equilibrium
temperature is generally much larger then the recoil
temperature. Thus, for the temperatures {\em below} the
Doppler temperature which we are aiming to achieve, the 
`stray' process {\em heats} the atoms, and {\em competes} with
the velocity selection induced by the two-photon 
transitions. Let us just
say here that it is possible, within some reasonable
assumptions, to write a Fokker-Planck equation
describing both one- and two-photon process (for
details, see ref. \cite{ref:vselect}), from which
the following conclusions can be drawn.

\begin{figure}
\begin{center}
  \epsfig{figure=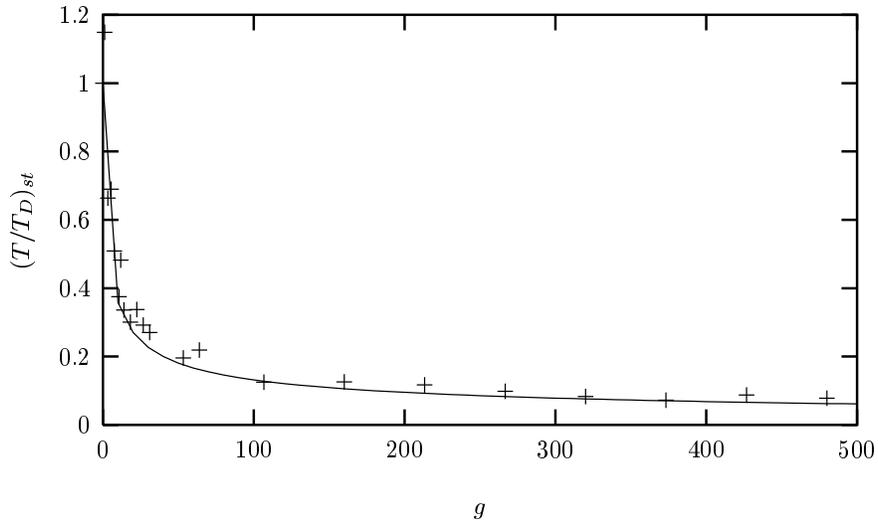,width=12cm,clip=}
\end{center}
\caption{Dependence of the temperature (normalized to the Doppler
temperature) as a function of the normalized two-photon transition
rate $g = W/(v/v_D)^2$, where the Doppler velocity is
$v_D = (k_B T_D/M)^{1/2} = (\hbar \Gamma/2M)^{1/2}$. In such units, typical
values of the recoil temperature range from 0.1 to 0.01 for usual
atomic species. The crosses correspond to a Monte-Carlo simulation.}
\label{fig:EqTemp}
\end{figure}

Fig.~\ref{fig:EqTemp} shows the equilibrium temperature as
a function of the two-photon transition rate. The
excited level is supposed to be a continuum so that
the two-photon transitions correspond to an irreversible 
ionization process. The irreversibility implies
that atoms are lost as they are ionized and that the
number of cooled atoms decreases along the selection
process. It is shown in ref. \cite{ref:vselect} that
the remaining ground-state population is
proportional to $(T/T_D)^{1/2}$. This brings about 
the discussion of the main limitation of the present
method, namely the fact that any generation of coherence by
state selection `looses' a fraction of atoms that
becomes larger as the selection becomes sharper.
There are (at least) two ways to prevent or limit this
undesirable effect, that we discuss now.

The first way of preventing depopulation is to inject
into the system `new' atoms in the ground state.
One can then achieve an
equilibrium regime in which the final selected state population
is constant. It is shown in ref.
\cite{ref:vselect} this scheme can produce, within experimentally
realistic conditions,
a significant {\em increasing of the phase space density}.

The second way to prevent depopulation is to `recycle'
the atoms lost in the selection process.
For example, one can take a discrete excited level in
Fig.~\ref{fig:levels} and let the atoms spontaneously
decay back to the ground state. However, in atomic systems
selection rules forbid the one-photon decay between states
coupled by two-photon transitions. The allowed
two-photon decay is considerably slower,
thus reducing the efficiency of the selection process.
A more convenient solution is to `fold' the level
scheme presented in Fig.~\ref{fig:levels} in such
a way that the excited level becomes in fact a
ground state hyperfine sublevel
(see Fig.~\ref{fig:RamanLevels}). We shall discuss
such a system in the next section.

\section{Raman cooling using quantum interference}
\label{sec:Raman}

\begin{figure}
\begin{center}
\epsfig{figure=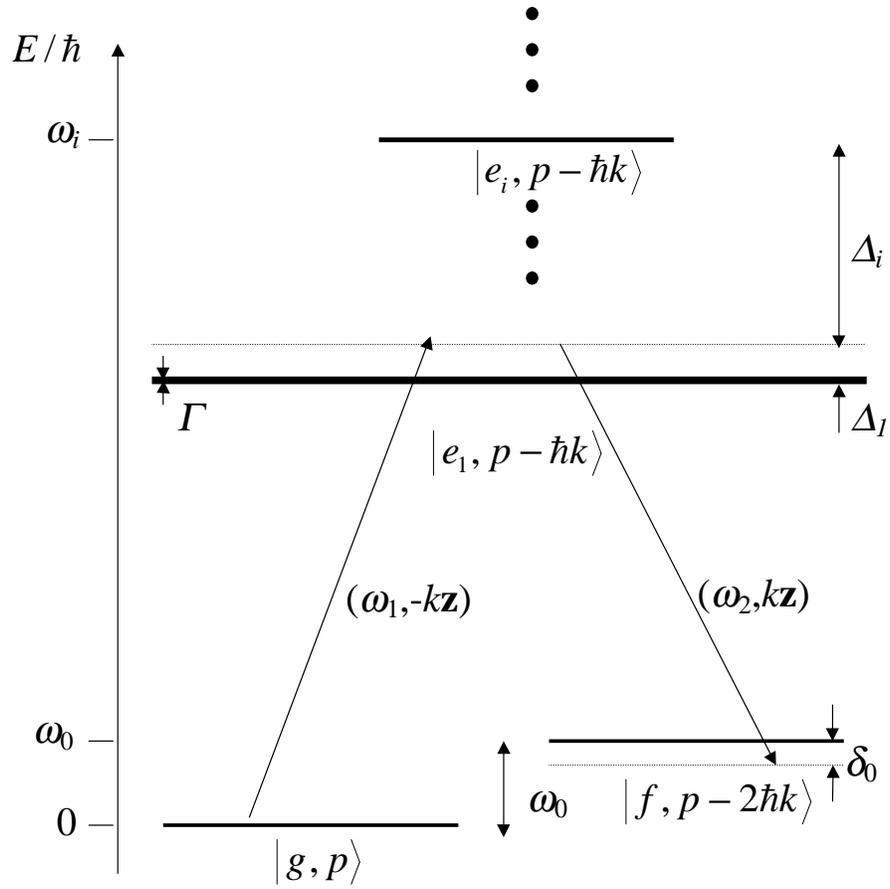,width=12cm,clip=}
\end{center}
\caption{
Levels involved in sub-recoil cooling by
Raman transitions. The total Raman transition rate is
proportional to the square of the sum of the contributions
from each intermediate level $|e_i\rangle$, leading to the
quantum interference effect.
  }
\label{fig:RamanLevels}
\end{figure}

Except in the subrecoil regime, the principle
of Raman cooling is quite similar to Doppler
cooling. A two-photon, Raman stimulated transition connects one
ground-state sublevel to another (Fig.~\ref{fig:RamanLevels}).
A laser beam is used
to excite a virtual intermediate level, and another beam,
whose frequency differs from the first one by a value very
close to the frequency interval between the two ground-state
sublevel, brings
the atoms to other substate by {\em stimulated} emission.
As spontaneous emission is not involved in the process,
the momentum transfer to the atoms
is perfectly controlled: if the two beams are counterpropagating,
the atomic momentum changes by $2 \hbar k$ in the process.
Cooling is performed by periodically exchanging
the direction of the beams (or by adding two extra Raman
beams \cite{ref:RamanInterf}). Choosing a negative value for
the Raman detuning $\delta_0$ favors the transition that cool the 
atoms \cite{ref:Raman} (exactly as in Doppler cooling).
However, once the atom arrives
in the second ground-state sublevel, it must
be bought back to first one in order to allow the cooling process
to proceed. This is done by `repumping'
the atoms with a beam resonant with the transition between
the second ground-state sublevel $|f\rangle$ and the excited
state. Once in the excited state, the atom has a chance of decaying
to the initial ground-state sublevel $|g\rangle$ (if the atom decays to
the sublevel $|f\rangle$, it will eventually be excited again, so
all atoms are finally `repumped' to  the sublevel $|g\rangle$).
As the repumping process randomizes the atomic velocity around the
recoil velocity, as described above, Raman cooling cannot attain subrecoil
temperatures.

To achieve the subrecoil regime, one should prevent cold enough
atoms from performing Raman transitions. An atom
of velocity $v$ sees an effective
Raman detuning shifted by $2kv$. Using  {\em pulsed} Raman excitation
whose spectrum is builded such that it presents a zero for $v=0$,
it is possible to decouple atoms
of low enough velocity from the Raman process: Atoms
within a certain range $\Delta v$ around $v=0$ are not excited. As
Raman transitions are very sharp, one can have
$\Delta v \ll v_r$. Once cooled to around the recoil velocity by
the Doppler-like process, the atoms perform random walks
in the velocity space until they fall into this
range, where they remain trapped, and the number of trapped atoms
increases as the cooling process goes on \cite{ref:Raman}.

\begin{figure}
\begin{center}
  \epsfig{figure=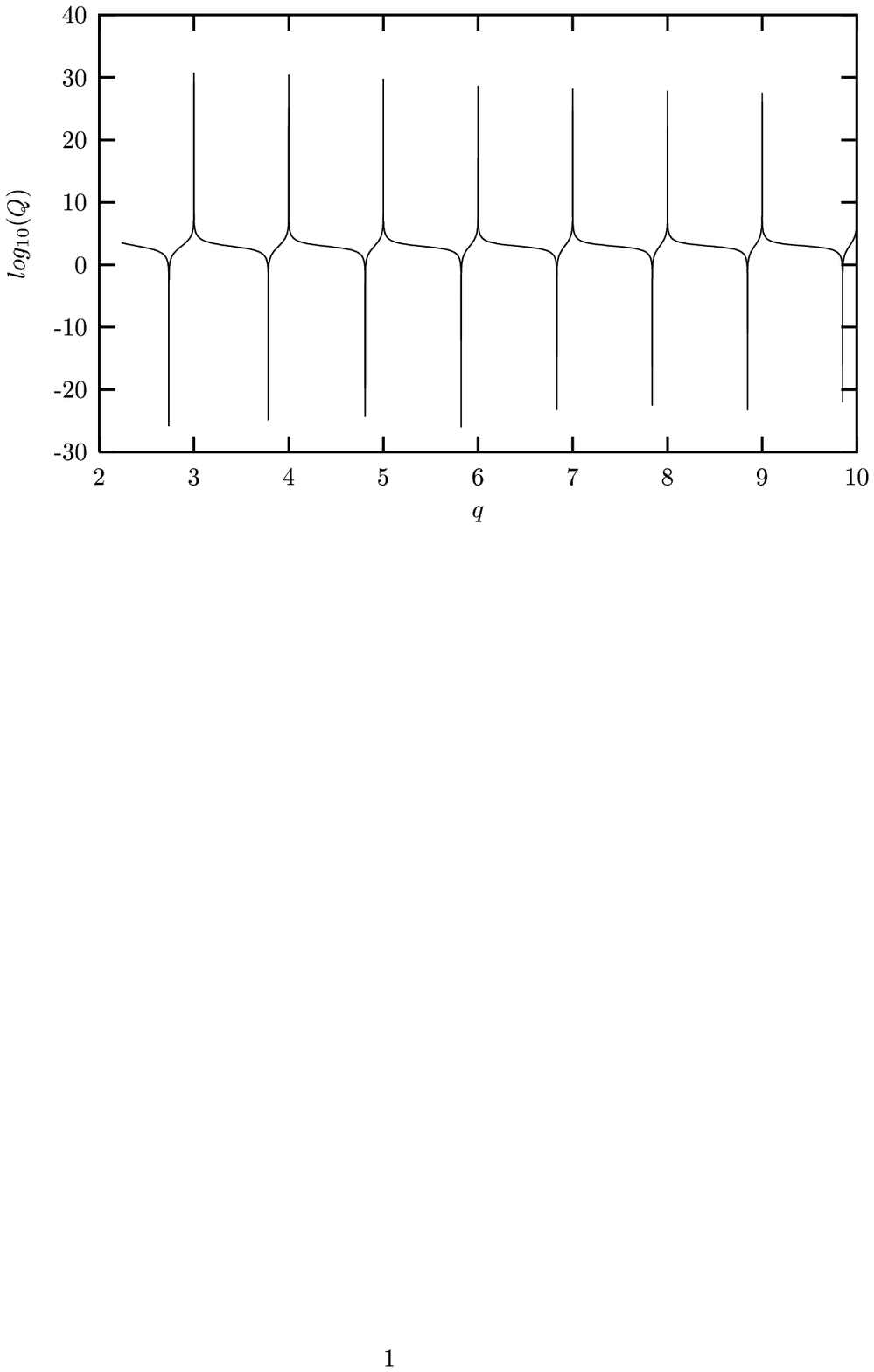,width=12cm,clip=}
\end{center}
\caption{Transition rate $Q$ for the stimulated
Raman transitions between the two ground-state sublevels
of the hydrogen atom. The parameter $q = \sqrt{\hbar \omega_1/Ry}$
coincides with an integer $n$ 
if the Raman excitation is resonant with the $1 \rightarrow n$
transition.}
\label{fig:Antiresonnances}
\end{figure}

We propose an alternative method that uses quantum interference instead
of spectrally shaped pulses to prevent subrecoil atoms from
being excited, thus allowing {\em continuous-wave} excitation.
Although hydrogen is {\em not}
the most interesting atomic species in the present context, we shall
consider here Raman transitions among the hyperfine sublevels
of the ground state of this atom, because exact expressions
for its dipole matrix elements are known, allowing the
calculation of the transition rate to be easily performed
\cite{ref:RamanInterf}.
Referring to  Fig.~\ref{fig:RamanLevels}, we can generalize
Eq. (\ref{eq:transrate}) for hydrogen
\begin{equation}
  Q = Q_0 \sum_n { A_n \over {1 \over q^2} - { 1 \over n^2 } }
  \label{eq:Htransrate}
\end{equation}
where we have used the fact that the hydrogen energy levels scales
as $1/n^2$ ($n$ integer) and have defined
$q \equiv [Ry/(\hbar \omega_1)]^{1/2}$
($Ry = 2.2 \times 10^{-18}$ J is the Rydberg). 
The Raman transitions amplitude is obtained by summing over all
possible intermediate levels $|e_i\rangle$. This means that
it can be possible, at least in particular cases, to
choose the frequency of the Raman beams in order
to produce {\em destructive} quantum interference among the
contributions of the intermediate levels.  The result of the
calculation outlined above is presented
in Fig.~\ref{fig:Antiresonnances} and shows a clear pattern of
resonances (corresponding to the usual resonances of the
hydrogen atom) and of `antiresonances' or `dark' resonances,
the latter corresponding to complete
destructive interference.
As Doppler effect displaces the position of the antiresonances,
an adequate tuning the Raman beams can prevent
subrecoil atoms from performing Raman transitions, thus decoupling
zero-velocity atoms from the Raman process, which allows
subrecoil cooling.

\begin{figure}
\begin{center}
  \epsfig{figure=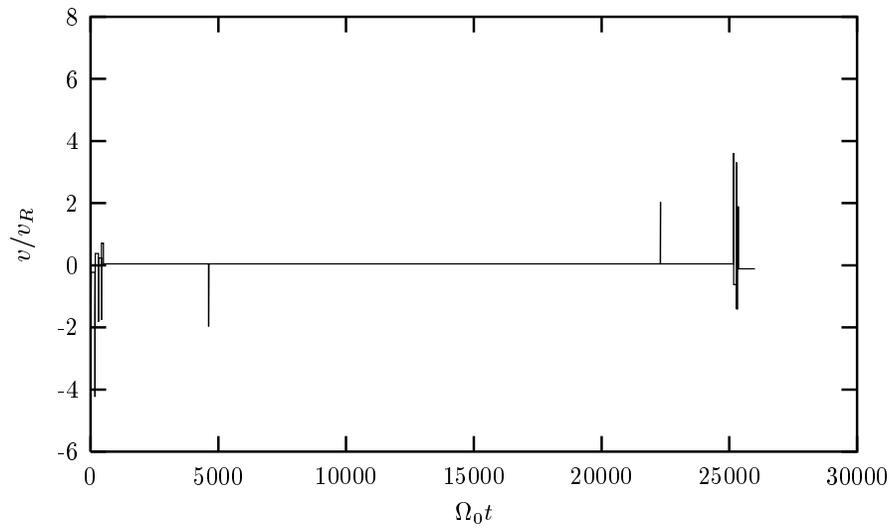,width=12cm,clip=}
\end{center}
\caption{Time evolution of the atom velocity under Raman cooling
  with quantum interference. The very long time the atom spends in
  a unique velocity state corresponds to a `Levy Flight'
  (for $5000 < \Omega_0t <22000$).}
\label{fig:LevyFlight}
\end{figure}

An interesting example of the particular dynamics of an atom under
subrecoil cooling with quantum interference is presented in Fig.
\ref{fig:LevyFlight}, showing the typical time evolution
of the atomic velocity.
It can be seen that once trapped in a low-velocity state,
the atom stays there for almost the entire duration of the
evolution. The statistics of the velocity
states is dominated by a single state:
This is the signature of a very special and
interesting behavior known as `Levy flights', which always
manifests itself in subrecoil cooling \cite{ref:LF}.

The present method allows {\em continuous-wave} Raman
subrecoil cooling, and, as the atoms can be recycled in the same way
as in the usual pulsed Raman cooling, it
performs subrecoil cooling without
populations losses. In other words, it performs {\em cooling}
instead of simply velocity selection, the repumping process
insuring the dissipation necessary to the cooling process.

\section{Conclusion}
\label{sec:Conclusion}

We have presented in a unified way a method for
generation of quantum coherence though controlled state selection
based on quantum interference. The application
of the method to concrete examples has been discussed,
concerning the generation of coherence among modes of
the electromagnetic field and atomic velocity selection,
in a variety of situations. Together
with many other applications of quantum interference
such as Velocity Selection by
Coherent Population Trapping and 
Electromagnetically Induced Transparency, 
the present method illustrates the power of quantum interference
as a tool for the manipulation of quantum entities.

\section{Acknowledgments}

The author thanks D. Hennequin, D. Wilkowski and
V. Zehnl\'e for fruitful discussions.
Laboratoire de Physique des Lasers, Atomes et
Mol{\'e}cules (PhLAM)
is {\it Unit\'e Mixte de Recherche} 8523 du Centre National de la
Recherche Scientifique (CNRS) et de l'Universit{\'e} des Sciences
et Technologies de Lille. Centre d'Etudes et Recherches
Lasers et Applications (CERLA) is supported by Minist\`{e}re de la
Recherche, R\'{e}gion Nord-Pas de Calais and Fonds Europ\'{e}en de
D\'{e}veloppement Economique des R\'{e}gions (FEDER).

\clearpage

\vspace{1cm}
\vspace{0.5cm}

\vspace{0.5cm}

\vspace{0.5cm}

\vspace{0.5cm}

\vspace{0.5cm}

\vspace{0.5cm}

\vspace{0.5cm}

\end{document}